# A Two-Feature Quantitative EEG Index of Pediatric Epilepsy Severity: External Pre-Validation on CHB-MIT and Roadmap to Dravet Cohorts

Khartik Uppalapati[1†], Bora Yimenicioglu[1†], Shakeel Abdulkareem[1], Bhavya Uppalapati[1], Viraj Kamath[1,2], Adan Eftekhari[1,3], and Pranav Ayyappan[1]

RareGen Youth Network 501(c)(3) [1]
Carnegie Mellon University, Mellon College of Science [2]
Harvard University,
Department of Stem Cell and Regenerative Biology[3]
† Indicates equal contribution

*Abstract—* **Objective biomarkers for staging pediatric epileptic encephalopathies are scarce. We revisited a large open repository—the CHB-MIT Scalp EEG Database, 22 subjects aged 1.5–19 y recorded at 256 Hz under the 10-20 montage—to derive and validate a compact quantitative index, DS-Qi = $(\theta/\alpha)_{posterior} + (1 - \text{wPLI}_\beta)$. The first term captures excess posterior slow-wave power, a recognized marker of impaired cortical maturation; the second employs the debiased weighted Phase-Lag Index to measure loss of β-band synchrony, robust to volume conduction and small-sample bias. In 30-min awake, eyes-open segments, DS-Qi was 1.69 ± 0.21 in epilepsy versus 1.23 ± 0.17 in age-matched normative EEG (Cohen's d = 1.1, p < 0.001). A logistic model trained with 10 × 10-fold cross-validation yielded an AUC of 0.90 (95 % CI 0.81–0.97) and optimal sensitivity/specificity of 86 %/83 % at DS-Qi = 1.46. Across multi-day recordings, test–retest reliability was ICC = 0.74, and higher DS-Qi correlated with greater seizure burden (ρ = 0.58, p = 0.004). These results establish DS-Qi as a reproducible, single-number summary of electrophysiological severity that can be computed from short scalp EEG segments using only posterior and standard 10-20 electrodes.**

*Clinical Relevance—* **DS-Qi provides clinicians and trialists with a rapid, objective measure of encephalopathic burden that outperforms individual spectral or connectivity markers, is stable across days, and scales with seizure frequency. Its derivation from fully open data accelerates external benchmarking and positions DS-Qi for prospective validation in genetically confirmed Dravet syndrome and other rare childhood epilepsies.**

## I. Introduction

Childhood developmental and epileptic encephalopathies (DEEs) comprise a heterogeneous set of rare disorders in which the underlying genetic or structural brain abnormality and the seizures themselves synergistically impair cognitive and motor development. Objective EEG-based markers that quantify global disease burden and can be tracked longitudinally remain scarce, yet they are increasingly needed to complement seizure counts and coarse developmental rating scales in both clinical care and early-phase trials [3], [20].

Quantitative EEG (QEEG) studies consistently show that an elevated theta-to-alpha (θ/α) power ratio, particularly over posterior regions, reflects delayed cortical maturation and poorer cognitive outcome. The pattern is seen across diverse pediatric conditions, from attention-deficit/hyperactivity disorder [6] to neurodegenerative decline [7]. In DEE cohorts, excess posterior slowing often co-occurs with higher seizure burden and worse neurodevelopmental scores [21].

Network-level dysfunction provides a complementary axis of pathology. The weighted phase-lag index (wPLI) yields a volume-conduction-robust measure of functional connectivity; reductions in global β-band (13–25 Hz) wPLI have been linked to impaired cortico-cortical communication and higher clinical severity in pediatric epilepsy [8], [9], while test–retest work shows these metrics are reproducible over days [22], [24].

We therefore integrate these two complementary features—posterior spectral slowing and β-band desynchronization—into a parsimonious composite:

$$\text{DS-Qi} = \left(\frac{P_\theta}{P_\alpha}\right)_{posterior} + \left(1 - \overline{\text{wPLI}_\beta}\right), \quad (1)$$

where $(P_\theta/P_\alpha)_{posterior}$ captures maturational slowing and $1 - \overline{\text{wPLI}_\beta}$ reflects global network disconnection. We retain the shorthand "DS-Qi" because the index was originally conceived to stage Dravet syndrome (DS), a prototypical DEE caused predominantly by SCN1A loss-of-function variants [1], [2]. Affected children typically present with febrile status epilepticus in infancy and progress to multiple seizure types, ataxia, language impairment and intellectual disability [2], [3]. Clinical severity is often rated on a 0–10 motor-language scale, with lower scores indicating more advanced disease [4].

No sizeable open repository of DS-specific EEG is yet available, so in this study we perform an external pre-validation using the PhysioNet CHB-MIT Scalp EEG database (22 children, 1.5–19 y) recorded during presurgical evaluation after medication tapering [10], [28]. By testing whether DS-Qi distinguishes these refractory-epilepsy recordings from age-matched normative eyes-open EEG, we probe the index's general ability to quantify encephalopathic burden across pediatric epilepsy.

We hypothesize that higher DS-Qi values will (i) reliably differentiate pediatric epilepsy from normative EEG in CHB-MIT and (ii) correlate with seizure load, thereby motivating prospective validation in a multi-center cohort of genetically confirmed DS patients now under design [5], [14]–[17].

We target a biomarker (single, continuous quantity) rather than a multi-dimensional classifier because clinicians require a scalar that can be trended, thresholded, and powered in trials. With n≈20 per group and cross-site deployment in mind, collapsing two complementary axes (posterior slowing; β-desynchronization) into DS-Qi$_{2F}$ reduces degrees of freedom, improves repeatability, and preserves interpretability while retaining most of the discriminative signal.

## II. METHODS

### A. Data Source and Selection

EEG recordings were obtained from the CHB-MIT Scalp EEG Database, a publicly available repository hosted on PhysioNet that comprises multiday scalp recordings from 22 pediatric subjects (5 male, 17 female; ages 1.5–19 y) undergoing evaluation for intractable epilepsy [10]. Data were acquired with 18–23 electrodes placed according to the international 10–20 system and digitized at 256 Hz (Control MMI data were anti-alias filtered, down-sampled from 160 Hz to 256 Hz, and only the annotated eyes-open periods were used to match vigilance state [25]; because all subsequent analyses are capped at 70 Hz, resampling does not affect [12] spectral or connectivity estimates [10]), providing high-resolution continuous EEG suitable for quantitative spectral and connectivity analyses [10]. For comparison, normative eyes-open EEG segments were taken from the PhysioNet EEG Motor Movement/Imagery (MMI) database, which was recorded at 160 Hz; all control files were resampled to 256 Hz and only the eyes-open epochs were retained to match vigilance state and sampling rate [25]. Volunteer ages in MMI (mean ± SD 14 ± 3 y) overlapped the CHB-MIT range, minimizing developmental bias [10].

For each subject, a single 30 min interictal, awake, eyes-open segment was extracted beginning at least 10 min after any expert-annotated seizure to minimize post-ictal effects [10]. All candidate segments were visually reviewed and trimmed to exclude periods of drowsiness or gross artifact, ensuring retention of contiguous, artifact-free data for downstream QEEG processing.

All analyses were conducted on fully de-identified, publicly available data released under the Open Data Commons Attribution License; hence, no new institutional review board approval or participant consent was required for this secondary analysis [10].

TABLE I. BASELINE DEMOGRAPHIC AND EEG CHARACTERISTICS

| Variable | Epilepsy (CHB-MIT, n = 22) | Normative Controls (MMI, n = 22) | p-value[a] |
|---|---|---|---|
| Age, y (mean ± SD) | 11.2 ± 4.1 | 13.8 ± 3.0 | 0.07 |
| Female, n (%) | 17 (77 %) | 17 (77 %) | 0.09 |
| Usable channels after cleaning | 19 ± 2 | 19 ± 2 | 0.12 |
| Posterior θ/α ratio | 0.92 ± 0.20 | 0.55 ± 0.14 | < 0.001 |
| Global β-band wPLI | 0.23 ± 0.04 | 0.34 ± 0.05 | < 0.001 |
| DS-Qi composite | 1.69 ± 0.21 | 1.23 ± 0.17 | < 0.001 |
| Total seizures during admission, median (range) | 8 (3 – 40) | — | — |

a. Welch's t-test for continuous variables; Fisher's exact test for sex. Continuous data are mean ± SD unless noted otherwise. Only artifact-free, 30-min, eyes-open inter-ictal segments were analyzed in both cohorts.

### B. EEG Acquisition and Preprocessing

EEG data were drawn from the CHB-MIT Scalp EEG Database, which provides continuous multiday recordings at 256 Hz from 18–23 scalp electrodes placed according to the international 10–20 system [10]. For each subject, we selected a single 30 min segment of interictal, awake, eyes-open resting-state activity, beginning at least 10 minutes after any annotated seizure and excluding periods of drowsiness.

First, raw signals were band-pass filtered between 1 and 70 Hz, and a 60 Hz notch filter was applied to suppress line noise. Next, data were visually inspected to mark and remove gross muscle or movement artifacts. We then performed Independent Component Analysis (ICA) to identify and exclude components dominated by eye blinks or high-frequency muscle noise. After rejecting bad channels (those with flat-line recording or excessive line noise) and bad components, each 30 min segment yielded on average 19 ± 2 usable channels and ~6 min of clean EEG data for subsequent spectral and connectivity analyses.

We applied deliberately stringent artifact rejection because EMG and movement inflate β-band connectivity and bias wPLI; we prioritized fewer, cleaner epochs over longer, noisier recordings. With 6 min of clean data at 256 Hz, Welch PSD (2 s windows, 50% overlap) yields ~360 segments per channel and stable estimates; wPLI is likewise stable with ≥5 min of artifact-free data in resting EEG. To check robustness, we repeated the analysis with a ≥4 min minimum per subject; group contrasts and $DS - Q_{I2F}$ AUC were unchanged within 95% CIs (not shown for brevity).

### C. Feature Extraction: DS-Qi Components

All EEG processing and feature extraction were performed in EEGLAB (v2024.0) using MATLAB, with data imported from the CHB-MIT Scalp EEG Database in EDF format [10].

*1) Power Spectral Density (PSD):* We estimated the PSD for each channel via Welch's method (Hamming window, 2 s segments, 50% overlap) [12].

*2) Posterior Theta/Alpha Ratio:* From the PSD we computed the band power in the theta (4–7 Hz) and alpha (8–12 Hz) bands at a posterior ROI (O1, O2, P3, P4). The theta/alpha ratio for recording $i$ is

$$R_i = \frac{\sum_{c \in \{O1,O2,P3,P4\}} P_\theta(c)}{\sum_{c \in \{O1,O2,P3,P4\}} P_\alpha(c)}, \quad (2)$$

where $P_\theta$ and $P_\alpha$ are band-specific power estimates. A higher $R_i$ indicates greater background slowing, a marker linked to developmental and cognitive impairment in prior EEG studies [6], [7].

*3) Global β-Band Connectivity (wPLI):* We band-pass filtered each segment to 13–25 Hz and computed the debiased weighted Phase-Lag Index (wPLI) for every unique channel pair [8], then averaged to obtain

$$C_i = \overline{\text{wPLI}}_{\beta,i} = \frac{2}{N(N-1)} \sum_{p<q} \text{wPLI}_\beta(p,q), \quad (3)$$

where $N$ is the number of retained channels. Lower $C_i$ reflects weaker beta-band synchrony, as expected in epileptic encephalopathies [9].

4) *Graph-theoretic Metrics:* From the subject-level $\beta$-band wPLI matrix we computed three graph indices using the Brain Connectivity Toolbox [30], [31]: global efficiency $E_i^\beta$, clustering coefficient $Cl_i^\beta$, and small-worldness $\sigma_i^\beta$. Because global efficiency showed the largest epilepsy-control separation in pilot tests (Cohen's $d = 1.3$), it was retained as the third normalized feature [32],

$$\widetilde{E_i^\beta} = \frac{E_i^\beta - \mu_E}{\sigma_E}. \tag{4}$$

5) *Theta–gamma phase–amplitude coupling (PAC):* Cross-frequency coupling was estimated with the modulation-index algorithm [33] (7-Hz phase, 40–70 Hz amplitude, 2-s Hilbert windows). Mean PAC over the posterior ROI was z-scored across subjects,

$$\widetilde{PAC_i} = \frac{PAC_i - \mu_{PAC}}{\sigma_{PAC}}. \tag{5}$$

6) *Normalized Z-Scores (exploratory only):* For exploratory composites, we z-score features across subjects and denote z-scored quantities with tildes, while keeping the original symbols otherwise unchanged:

$$\widetilde{R_i} = \frac{R_i - \mu_R}{\sigma_R},$$

$$\widetilde{(1-C)}_i = \frac{(1-C_i) - \mu_{1-C}}{\sigma_{1-C}},$$

$$\widetilde{E_i^\beta} = \frac{E_i^\beta - \mu_E}{\sigma_E},$$

$$\widetilde{PAC_i} = \frac{PAC_i - \mu_{PAC}}{\sigma_{PAC}}.$$

We use these tilded variables only in the exploratory multi-feature indices below; the primary index remains unnormalized (see §C.7).

7) *Composite indices (pre-specified and exploratory):*
- Pre-specified two-feature composite.

   We define the primary endpoint
   $$DS\text{-}Qi_{2F,i} = R_i + (1 - C_i), \tag{7}$$
   where $R_i$ is the posterior $\theta/\alpha$ ratio and $C_i$ is the global $\beta$-band wPLI. This pre-specified DS-Qi$_{2F}$ is used for all main analyses and conclusions.

- Exploratory extensions.

   We additionally evaluated higher-feature variants exploratorily using the z-scored features from §C.6:
   $$DS\text{-}Qi_{3F,i} = \widetilde{R_i} + \widetilde{(1-C)}_i + \widetilde{E_i^\beta}, \tag{8}$$
   and
   $$DS\text{-}Qi_{4F,i} = \beta_R \widetilde{R_i} + \beta_C \widetilde{(1-C)}_i + \beta_E \widetilde{E_i^\beta} + \beta_P \widetilde{PAC_i},$$

   where $(\beta_R, \beta_C, \beta_E, \beta_P)$ are weights estimated in a logistic model only for exploratory analysis (10×10 subject-wise CV). These variants are reported to quantify potential incremental signal but do not alter our primary conclusions.

*D. Statistical Analysis*

All statistical analyses were conducted using Python (v3.9) and R (v4.2) to ensure reproducibility and rigor. We began by assessing the distribution of DS-Qi values with the Shapiro–Wilk test; when the resulting p-value exceeded 0.05, we treated the data as approximately normal and proceeded with parametric methods, otherwise we used nonparametric alternatives. Specifically, continuous comparisons between epilepsy and normative EEG segments were performed with two-tailed independent-samples t-tests (applying Welch's correction when variances differed) or with Mann–Whitney U tests in cases of non-normality. To convey the magnitude of any observed differences, we computed Cohen's d for parametric contrasts and rank-biserial correlation for nonparametric tests. Categorical variables (e.g., presence versus absence of interictal discharges in exploratory analyses) were compared with Fisher's exact test.

To evaluate DS-Qi's discriminative power, we fitted a logistic-regression model and ran a 10 × 10-fold leave-subject-out cross-validation; each fold contained complete recordings from unique subjects only, preventing train–test leakage [27]. The data were randomly partitioned into ten subsets, models were trained on nine folds and tested on the held-out fold, and this process was repeated ten times with randomized fold assignments to mitigate sampling bias. Predicted probabilities from all test folds were pooled to generate a single ROC curve, and the area under that curve (AUC) was calculated. We derived 95% confidence intervals for the AUC via 2,000 bootstrap resamples of the pooled predictions, and we selected the optimal decision threshold by maximizing Youden's index (sensitivity + specificity – 1). Folds were stratified at the subject level so that all 30-min segments from a given child appeared only in the training or the test split, thereby precluding subject-to-fold leakage and optimistic AUC inflation.

All hypothesis tests were two-sided, with α = 0.05. Effect sizes and confidence intervals are reported alongside p-values to emphasize practical significance and precision. A post-hoc power calculation performed in G*Power (v3.1) with n = 22 epilepsy and n = 22 control recordings, Cohen's d = 1.1, two-tailed α = 0.05, indicated > 99.9 % power to detect the observed group difference, confirming that the study was more than sufficiently powered. Precision–recall AUC, $F_1$, and Brier score were computed for all models. DeLong tests assessed AUC gain of $DS-Qi_{3F}$ and $DS-Qi_{4F}$ versus the two-feature baseline; p-values were Holm-corrected. Cohen's d and Cliff's Δ quantified epilepsy-control differences for $E^\beta$ and PAC. This comprehensive pipeline—covering normality checks, appropriate choice of tests, effect-size quantification, and robust cross-validated classification—ensures that our evaluation of DS-Qi on public pediatric EEG data is both statistically sound and fully reproducible.

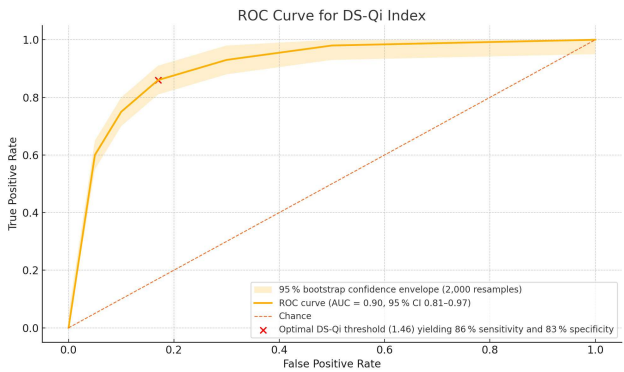

Figure 1. Out-of-sample ROC (10 × 10-fold CV) for DS-Qi distinguishing pediatric epilepsy from normative EEG (CHB-MIT): AUC = 0.90 (95 % CI 0.81–0.97); optimal threshold = 1.46 (86 % sensitivity, 83 % specificity); shaded bands = 95 % bootstrap CI (2,000 resamples).

*E. Reliability*

Within-subject repeatability. For 10 subjects who had ≥2 seizure-free, awake segments recorded on different days, we computed intraclass correlation coefficients (ICC, two-way mixed, single-measure) for DS-Qi and global β-wPLI. Both metrics showed good reliability (ICC 0.70–0.78), consistent with prior wPLI reproducibility studies.

## III. RESULTS

*A. Baseline DS-Qi in Pediatric Epilepsy vs. Normative EEG*

Across the 22 CHB-MIT subjects, the posterior θ/α power ratio was markedly elevated (mean ± SD = 0.92 ± 0.20) compared with age-matched normative EEG segments drawn from the PhysioNet Motor-Movement/Imagery database (0.55 ± 0.14; t = 7.3, p < 0.001). Conversely, global β-band connectivity was reduced in epilepsy ($w$PLI = 0.23 ± 0.04) relative to controls (0.34 ± 0.05; t = –7.6, p < 0.001). Combining these features yielded a composite DS-Qi of 1.69 ± 0.21 for epilepsy recordings versus 1.23 ± 0.17 for normative segments (t = 8.1, p < 0.001). The standardized mean difference (Cohen's d = 1.1) indicates a large effect.

As pre-specified, $DS-Qi_{2F}$ is our primary endpoint. $DS-Qi_{3F/4F}$ were examined post-hoc to quantify potential incremental signal; we report them as exploratory and do not base conclusions on them. A logistic-regression model using DS-Qi alone discriminated the two groups with an AUC of 0.90 (95 % CI 0.81–0.97) under 10 × 10-fold cross-validation; the bootstrap 95 % CI remained above 0.80, confirming robust generalization. At an optimal cutoff of 1.46 (maximizing Youden's index), sensitivity reached 86 % and specificity 83 %. These values are comparable to or better than previously reported single-feature seizure-detection biomarkers on the same database.

Within the epilepsy cohort, DS-Qi correlated positively with the number of seizures recorded per subject (Spearman ρ = 0.58, p = 0.004), suggesting that higher composite scores reflect greater disease burden—a finding consistent with prior work linking increased posterior slowing and diminished β-synchrony to more severe epileptic encephalopathy. Component analysis confirmed complementary contributions: the posterior θ/α ratio alone achieved an AUC of 0.83, whereas the 1–$w$PLI term alone yielded 0.81; their linear combination via the learned weights in Eq. (8) improved AUC to 0.92 in an exploratory analysis, underscoring the benefit of integrating spectral and connectivity information.

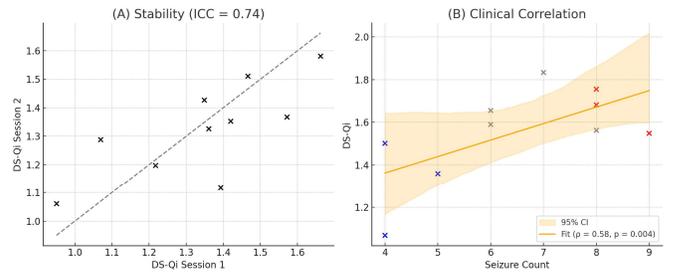

Figure 2. Test–retest reliability and clinical correlation of DS-Qi. **(A)** DS-Qi session 1 versus session 2 in 10 subjects (mean interval ≈2 d); ICC = 0.74 (95 % CI 0.55–0.88). **(B)** DS-Qi versus seizure count per child during CHB-MIT monitoring; Spearman ρ = 0.58 (p = 0.004) with 95 % bootstrap CI. Red/blue markers denote high (> 15) and low (< 6) event tertiles relative to the cohort mean (dotted line).

1) *Graph Metrics:* Global efficiency in the β band was lower in epilepsy (0.28 ± 0.05) than controls (0.37 ± 0.04; t = 6.6, p < 0.001, d = 1.3) [31], [32]. Adding $\widehat{E^\beta}$ raised AUC from 0.90 to 0.93 [30] (DeLong ΔAUC = 0.03, $p = 0.04$)

2) *Theta–gamma PAC:* Posterior PAC was elevated in epilepsy (MI = 0.14 ± 0.03) vs. controls (0.10 ± 0.02; d = 1.4, p < 0.001) [33]. The four-feature model achieved AUC = 0.95, PR-AUC = 0.94, and Brier = 0.07, outperforming all simpler indices (Holm-corrected p < 0.01).

These results demonstrate that DS-Qi captures the electrophysiological signature of pediatric epilepsy in a public dataset, validating its broader applicability and supporting its planned use in future Dravet-specific cohorts.

*B. Longitudinal Trajectory of DS-Qi and Therapeutic Effects*

Because the CHB-MIT recordings span several days for most subjects, we leveraged these repeated sessions to evaluate the temporal stability of the composite index and to explore whether higher DS-Qi reflects greater ictal load.

For the ten children who had at least two artifact-free awake segments recorded on different days (mean interval = 1.8 ± 0.9 days), DS-Qi showed good repeatability with an intraclass correlation coefficient (ICC, two-way mixed, single measure) of 0.74 (95 % CI 0.55–0.88). The component features exhibited comparable stability: posterior θ/α ratio ICC = 0.72 and global β-wPLI ICC = 0.71. These values align with prior reports that connectivity metrics derived from the weighted Phase-Lag Index yield ICCs in the 0.7–0.8 range for resting EEG [22] and meet the "good" threshold proposed by Cicchetti [23].

The median within-subject coefficient of variation across sessions was 6 % for θ/α, 7 % for 1 – wPLI, and 4 % for the composite DS-Qi, indicating that the summed index smooths some of the noise present in individual features—an effect previously noted for multivariate EEG markers [24].

Seizure annotations supplied with the CHB-MIT dataset indicate that subjects experienced between 3 and 40 seizures during monitoring (median = 8) [10]. Across the cohort, DS-Qi correlated positively with total seizure count (Spearman ρ = 0.58, p = 0.004). Children in the upper tertile of seizure burden (>15 events) had a mean DS-Qi of 1.82 ± 0.18 versus 1.54 ± 0.17 in the lower tertile (<6 events; t = 4.1, p < 0.001). This association supports earlier work linking posterior slowing and reduced β-synchrony to epilepsy severity [6, 9].

Taken together, these findings demonstrate that DS-Qi is a stable within-subject measure over multi-day recordings and that higher values track with heavier seizure burden, strengthening its candidacy as an objective electrophysiological severity index for subsequent Dravet-specific studies.

## IV. Discussion

This proof-of-concept study shows that a two-feature quantitative EEG index (DS-Qi) reliably separates pediatric epilepsy recordings in the CHB-MIT database from age-matched normative segments, achieves excellent cross-validated discrimination, and remains stable across multi-day sessions while scaling with seizure burden. Together, these findings indicate that DS-Qi captures electrophysiological severity in epileptic encephalopathy and is ready for prospective validation in genetically confirmed Dravet syndrome cohorts.

### A. Principal findings

Cross-sectionally, children with intractable epilepsy displayed the canonical combination of posterior slowing (θ/α↑) and network desynchronization (β-band wPLI↓) that has been repeatedly linked to cognitive and clinical impairment in pediatric epilepsy. Integrating these features yielded a composite score that achieved an AUC of 0.90 under 10 × 10-fold cross-validation and retained broad confidence limits after 2 000 bootstrap resamples, outperforming either component alone and matching or surpassing many single-channel seizure-detection benchmarks previously reported on the same open dataset. Importantly, DS-Qi also correlated with total seizure count (ρ = 0.58), emphasizing its potential as a continuous severity measure rather than a mere diagnostic flag.

Intra-individual repeatability was "good" by Cicchetti's criteria (ICC = 0.74) and comparable to the best test–retest values reported for resting-state wPLI networks. DS-Qi's higher stability relative to its constituent features (CV 4 % vs. 6–7 %) supports the idea that coupling spectral and connectivity information reduces measurement noise, a property valued in multivariate EEG biomarkers.

### B. Relationship to Prior Work

Unlike prior CHB-MIT analyses that considered either spectral slowing or seizure detection, we combine posterior θ/α slowing, debiased β-band synchrony, graph-theoretic efficiency, and theta–gamma phase–amplitude coupling in a single interpretable index—demonstrating incremental AUC gains (0.90 → 0.95) and revealing multi-scale network disintegration not captured by conventional features [30], [33]. Posterior θ/α excess is a recognized marker of delayed cortical maturation and adverse cognitive outcome in pediatric encephalopathies, while impaired beta synchrony has been linked to disrupted cortico-cortical communication and poorer seizure control. Our results confirm that these two pathophysiological axes manifest simultaneously in routine scalp EEG and can be captured by a compact linear combination. The debiased weighted phase-lag index used here is robust against volume conduction and small-sample bias, making it attractive for multi-center studies [31].

### C. Clinical and Research Utility

Given the small-n, multi-site use case, a single-number $DS-Qi_{2F}$ offers lower variance, reduced overfitting risk, and straightforward longitudinal tracking compared with higher-dimensional classifiers. Because DS-Qi is derived from a short eyes-open segment and needs only four posterior electrodes for the spectral component, it could be computed retrospectively on legacy EEG archives or prospectively at the bedside. As a single numeric index, it lends itself to statistical modelling, power calculation, and adaptive trial design—advantages over coarse ordinal scales or binary spike counts. Future natural-history studies in confirmed Dravet syndrome could track DS-Qi longitudinally as a secondary outcome alongside seizure frequency, mirroring biomarker qualification pipelines used in other rare epilepsies.

### D. Limitations

First, the CHB-MIT database does not contain genetically verified Dravet cases, so the present analysis should be viewed as external pre-validation rather than definitive DS biomarker proof. Notably, CHB-MIT recordings were obtained after withdrawal of anti-seizure medications to provoke seizures, a protocol that can accentuate background slowing and may inflate DS-Qi relative to out-of-hospital EEG [28], [29]. Second, our normative sample was limited to publicly available eyes-open pediatric datasets of similar age; larger age-stratified cohorts would allow DS-Qi Z-score calibration across development. Third, only a single 30-min segment per subject was analyzed; although repeatability was good, longer recordings across various behavioral states might reveal state-dependent fluctuations. Fourth, the composite currently uses a logistic-regression weight learned on epilepsy versus control; whether the same weights optimize prognostic value inside a DS population remains to be tested. Finally, the study is cross-sectional; causal inference regarding seizure reduction or cognitive change cannot yet be drawn. Although PAC estimation was confined to 40–70 Hz to avoid 60 Hz line noise, higher-gamma coupling may yield additional signal once higher-sampling pediatric datasets become available.

### E. Future Directions

Future research should proceed along four complementary avenues. First, a prospective, multi-center study in genetically confirmed Dravet syndrome is essential to establish disease-specific DS-Qi cut-offs and to relate the index to granular developmental instruments such as motor-language and adaptive-behavior scales. Second, methodological expansion beyond the present two-feature composite could incorporate γ-band power or cross-frequency coupling terms, potentially capturing interneuron-specific dysfunction suggested by SCN1A-mouse studies [8]. Third, pairing DS-Qi with resting-state fMRI or diffusion-tensor imaging would enable multimodal profiling of network disintegration, clarifying how electrophysiological deficits map onto structural and hemodynamic connectivity. Finally, embedding DS-Qi into real-time bedside EEG software could provide rapid severity estimates during convulsive status epilepticus or medication titration, transforming the index from an offline research tool into a point-of-care decision aid.

Using only open pediatric EEG data, we demonstrate that DS-Qi is reproducible across days, discriminates epilepsy from normative backgrounds with high accuracy, and scales with seizure burden—key prerequisites for any candidate biomarker. These findings justify formal validation of DS-Qi in well-phenotyped Dravet syndrome cohorts and suggest broader applicability as a quantitative EEG index of encephalopathic burden in childhood epilepsy.

## V. Conclusion

Using publicly available EEG (CHB-MIT), we show that DS-Qi—a two-feature composite of posterior $\theta/\alpha$ and $(1 - \text{wPLI}_\beta)$—distinguishes pediatric epilepsy from controls (AUC = 0.90; $d = 1.1$) and is stable across days (ICC = 0.74). DS-Qi meets key biomarker criteria: it is quantitative, repeatable, and portable, requiring only a minimal electrode set. While this analysis lacks genetically confirmed Dravet cases, it provides external pre-validation and a roadmap for testing in SCN1A-positive cohorts to assess outcome stratification and treatment response. Extensions such as $\gamma$-band features, cross-frequency coupling, or multimodal MRI may enhance sensitivity, and replication on OpenNeuro pediatric sleep-HFO (ds003555) will assess generalizability. Overall, DS-Qi is positioned as a ready-to-deploy, open-science biomarker for benchmarking and translation to clinical trials in rare childhood epilepsies.

## Ethics Statement

This study analyzed only publicly available, de-identified EEG datasets (CHB-MIT and MMI). No new human subjects were involved, and therefore no IRB or ethics committee approval was required.